\documentclass[letterpaper]{jpconf}
\usepackage{graphicx}
\begin{document}
\title{Reviewing the Evidence for Pentaquarks}

\author{Alex R. Dzierba\footnote{Presented this paper at the First Meeting of the APS Topical Group on 
Hadronic Physics}}

\address{Department of Physics, Indiana University, Bloomington, IN 47405}

\ead{dzierba@indiana.edu}

\author{Curtis A. Meyer}

\address{Department of Physics, Carnegie Mellon University, Pittsburgh, PA 15213}

\ead{cmeyer@ernest.phys.cmu.edu}

\author{Adam P. Szczepaniak}

\address{Nuclear Theory Center, Indiana University, Bloomington, IN 47405}

\ead{aszczepa@indiana.edu}

\begin{abstract}
Several experimental groups have reported evidence for baryons with 
flavor exotic quantum numbers that cannot be explained as $qqq$ bound states
but require a minimum of five quarks -- $qqqq \bar q$.  These pentaquark states 
include the $\theta^{+}$, the $\Xi^{--}$ and the $\theta_{c}$.   The reported 
widths of these new states are consistent with experimental resolution and may 
be as narrow as a few MeV/$c^2$ or less.   Prior to 2003, experimental searches 
for flavor exotic baryons spanning several decades yielded negative results. 
There have also been a number of searches carried out since the reports of these 
new pentaquark states that do not confirm their existence.  This review of both 
the positive and negative reports seeks to understand the current 
situation regarding the experimental evidence for pentaquarks.
\end{abstract}.

\section{Introduction}
It is noteworthy that except for one or two possible exceptions in the
meson sector,  the hundreds of known hadrons can be described as
bound states of three quarks ($qqq$), in the case of baryons, or a
quark and anti-quark ($q \bar q$), in the case of mesons.  There have
been reports of possible evidence of mesons with exotic $J^{PC}$ quantum
numbers, not possible for $q \bar q$, but until 2003,  no reports of
baryons with quantum numbers inconsistent with $qqq$.  There were
searches in the 1960's and 1970's for what was then referred to as the $Z^*$ -- a baryon
with positive strangeness. The review by
Hey and Kelly \cite{hey_kelly} discusses these early searches,
mostly in bubble chamber experiments,  and
a more recent review by Trilling  of the current exotic baryon sightings \cite{pdg} lists publications
reporting on bubble chamber experiments with incident
$\pi$ and  $K$ beams searching for the  $Z^*$.
Those experiments found no enhancements in $S=+1$ baryon channels.
  Indeed the failure to find such flavor exotic
baryons in the 1960's and 1970's
 lent credence to the then nascent quark model.  
 
\begin{figure}[h]
\includegraphics[width=18pc]{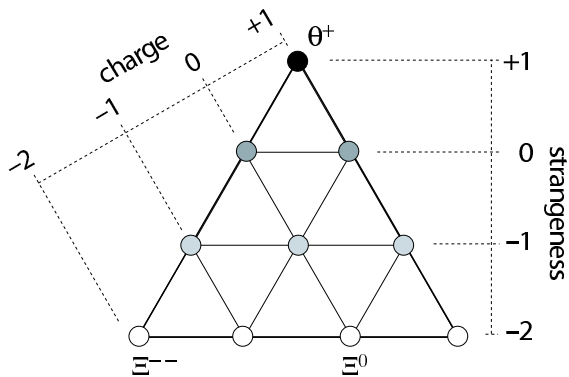}\hspace{2pc}%
\begin{minipage}[b]{18pc}\caption{\label{pentagram}The predicted 
anti-decuplet \cite{diakonov} of pentaquark
baryons indicating, in particular,
the states $\theta^+ = uudds$, $\Xi^{--}_5=ddss \bar u$ and
$\Xi^0_5=udss\bar d$.  Evidence for these states have been presented
as well as for a $\theta_c = uudd \bar c$.  Other searches for these
states have yielded null results.}
\end{minipage}
\end{figure}

 In 1997, 
D.~Diakonov, V.~Petrov and M.~Polyakov \cite{diakonov},
using a chiral soliton model, predicted a $J^{P}=\frac{1}{2}^{+}$ 
exotic anti-decuplet of pentaquark baryons (see Figure~\ref{pentagram})
including a $uudd \bar s$ baryon with positive
strangeness, isospin zero and a mass of 1530~MeV/$c^2$
with a width of 15~MeV/$c^2$.  Evidence for that state, the $\theta^+$,
has been reported by several experiments.
That anti-decuplet is also predicted to include a  $\Xi^{--}_5=ddss \bar u$ and a
$\Xi^0_5=udss\bar d$ for which evidence has also been reported.
Yet another experiment claims an anti-charm
 pentaquark baryon,  $\theta_c = uudd \bar c$.  These findings have
 generated much excitement in the nuclear and particle physics
 theory community -- nearly 300 papers have been written on the
 interpretation of these states and their properties.
  But as will be discussed
 below, these purported discoveries have also inspired recent searches
that  fail to find evidence for
 these states.

\section{Experiments claiming positive pentaquark signals}

Table~\ref{posex} lists the experiments claiming evidence for pentaquark states.
There are eleven  experiments claiming a $\theta^+$, a
 state with $S=+1$, a mass of
about 1.54~GeV/$c^2$ and a width consistent with being less than
 the mass resolutions of the the experiments.  The state is observed through
 its decay into either $K^+n$ or $K_S^0p$.  The first five reactions listed in
 Table~\ref{posex} used a photon probe of relatively low energy, a few GeV,
 and the purported pentaquark is observed in the $K^+n$ mode.
 The first sighting of the  $\theta^+$ was from the LEPS experiment at
 Spring8 \cite{spring8,spring8-2} followed shortly thereafter by the CLAS experiment
 at Jefferson Lab \cite{clas1,clas2}.   The SAPHIR experiment is also a low-energy photon experiment \cite{saphir}.
 
 \begin{table}[h]
\caption{\label{posex}Positive signals for pentaquark states.  Please see the
text regarding the final state neutron in the LEPS, CLAS and SAPHIR experiments.}
\begin{center}
\begin{tabular}{llllcl}
\br
Experiment&Reaction&State&Mode&Reference\\
\mr
LEPS(1)&$\gamma C_{12} \to K^+K^-X$&$\theta^+$&$K^+n$&\cite{spring8}\\
LEPS(2)&$\gamma d \to K^+K^-X$&$\theta^+$&$K^+n$&\cite{spring8-2}\\
CLAS(d)&$\gamma d \to K^+K^-(n)p$&$\theta^+$&$K^+n$&\cite{clas1}\\
CLAS(p)&$\gamma p \to K^+K^-\pi^+(n)$&$\theta^+$&$K^+n$ &\cite{clas2}\\
SAPHIR&$\gamma p \to K^0_SK^+(n)$&$\theta^+$&$K^+n$ &\cite{saphir}\\
COSY&$p p \to  \Sigma^+K^0_S p$&$\theta^+$&$K_S^0p$ &\cite{cosy}\\
JINR&$p(C_3H_8)  \to K_S^0 pX$&$\theta^+$&$K_S^0p$&\cite{jinr}\\
SVD&$p A \to K_S^0 pX$&$\theta^+$&$K_S^0p$ &\cite{svd}\\
DIANA&$K^+  Xe \to K_S^0 p (Xe)^{\prime}$&$\theta^+$&$K_S^0p$ &\cite{diana}\\
$\nu$BC &$\nu A \to K_S^0 pX$&$\theta^+$&$K_S^0p$ &\cite{bebc}\\
NOMAD &$\nu A \to K_S^0 pX$&$\theta^+$&$K_S^0p$ &\cite{nomad}\\
HERMES&quasi-real photoproduction&$\theta^+$ &$K_S^0p$&\cite{hermes}\\
ZEUS&$\e p \to K_S^0 pX$&$\theta^+$&$K_S^0p$&\cite{zeus}\\
\mr
NA49&$p p \to \Xi \pi X$&$\Xi_5$&$\Xi \pi$&\cite{na49}\\
\mr
H1&$e p \to (D^*p)X$&$\theta_c$&$D^*p$&\cite{h1}\\
\br
\end{tabular}
\end{center}
\end{table}
 
 The final state neutron in the above five experiments is not detected.  In the LEPS
 experiment the assumed $K^+n$ effective mass is actually the missing mass
 recoiling against the $K^-$.  In the CLAS(d) experiment the assumption is 
 that the final state proton, which is detected, is the spectator nucleon
 and the $K^+n$ effective mass is the mass recoiling against 
 the $K^-p$ system.  In
 the CLAS(p) experiment the neutron is inferred from missing mass and in the
 SAPHIR experiment the neutron is inferred from kinematic fitting.

The COSY experiment used a low-momentum proton beam \cite{cosy} spanning
 the momentum range from 2.85 to 3.3~GeV/$c$.  The JINR result comes from an analysis
 of collisions in a propane bubble chamber exposed to a 10~GeV/$c$ proton beam.
 The SVD experiment \cite{svd} at IHEP studied $pA$ collisions at 70~GeV/$c$.  Both the
 DIANA and $\nu$BC groups re-analyzed old data from a liquid xenon bubble chamber
 (in the case of DIANA) \cite{diana} and the CERN BEBC and the FNAL 15-foot chamber
 (in the case of $\nu$BC) \cite{bebc} -- the former using a low-energy $K^+$ beam, the 
 the latter neutrinos.  The NOMAD experiment at CERN was built to search for neutrino
 oscillations \cite{nomad}.  The HERMES experiment at DESY found evidence for
 the $\theta^+$ with quasi-real photons \cite{hermes} and ZEUS claims evidence \cite{zeus} for
 the $\theta^+$ in $ep$ collisions.  The evidence for the the $\theta^+$ presented
 in references \cite{spring8,spring8-2,clas1,clas2,saphir} is in the $K^+n$ mode, which is manifestly
 flavor-exotic while the other experiments report the $\theta^+$ in the $K_S^0p$ mode, which
 is a linear combination of $S=+1$ and $S=-1$.  Several of these experiments 
 studied the $K+p$ spectrum and found no evidence for a $\theta^{++}$, thus 
 concluding that $I=1/2$ for the $\theta^+$.
 
 At first glance, the evidence for the  $\theta^+$ seems convincing.  The
 experiments claim signals with a statistical significance ranging from 4 to 7
 standard deviations over background (we will return to this issue later).
 The evidence comes from low-energy and high energy experiments and
 experiments with a variety of beams: photons (real and quasi-real), 
electrons, protons, neutrinos and positively charged kaons.

Evidence for the pentaquark cascade states, $\Xi^{--}_5$ and $\Xi^0_5$,
comes from a single experiment, NA49 at CERN \cite{na49}, in the 
$\Xi^- \pi^-$ and $\Xi^- \pi^+$ modes in proton-proton collisions
at $\sqrt{s}=17.2$~GeV.   The reported mass is 1.862~GeV/$c^2$
and the width is consistent with being below detector resolution. And evidence
for the anti-charmed pentaquark, $\theta_c$,  also comes from a 
single experiment, H1 at HERA \cite{h1}, in the $D^{*+} \bar p$
and $D^{*-}  p$ modes at a mass of 3.1~GeV/$c^2$ and a width
consistent with detector resolution.  The data were collected in $ep$
collisions at $\sqrt{s}$ of 300 and 320~GeV. 

\section{Pentaquark searches with null results}

Table~\ref{negex} lists recent experiments that have searched for
and failed to find evidence for  pentaquark
signals.  As discussed earlier  \cite{hey_kelly,pdg},
searches for $S=+1$ baryons in the 1960's and 1970's  also failed to find evidence for their existence.

The experiments in Table~\ref{negex} are listed in alphabetical order by name.
For each of the three pentaquark states ($\theta^+$, $\Xi_5$ and $\theta_c$), we
indicate with a $\Uparrow$ that the state was observed or with a $\Downarrow$
that the state was searched for and not observed.  The entry ($-$) indicates that
the state was not searched for.  

Of the 18 entries in Table~\ref{negex}, 16 searches for the $\theta^+$ yielded
null results.  One experiment, WA89 \cite{wa89}, that failed to find the $\Xi_5$,
did not search for the  $\theta^+$
and another experiment, ZEUS, that searched for and failed to find either the 
$\Xi_5$ or $\theta_c$ did claim positive evidence for the $\theta^+$ \cite{zeus,cascade,zeus_charm}.
All of the experiments (9 of them) searching for the $\Xi_5$ failed to find the state
and all of the experiments (6 of them) searching for the $\theta_c$ failed to
find that state.  As will be discussed below, the experiments listed in
Table~\ref{negex} are characterized by high statistics and excellent mass
resolution, of order 1 to 2~MeV/$c^2$ in most cases.  These experiments also
see the relevant {\it benchmark} states, such as the $\phi(1020)$, $K^*(890)$, $D^*$,
$\Lambda(1520)$, $\Xi(1320)$ or $\Xi(1530)$, with statistics that overwhelm experiments that
report positive pentaquark sightings.

\begin{table}[h]
\caption{\label{negex}Recent negative searches for pentaquark states.  For each 
pentaquark state ($P$) we indicated with a $-$ that the state was not included in the
search while $\Downarrow$ indicates that the state was searched for and not
observed and $\Uparrow$ indicates that the state was searched for and
observed.}
\begin{center}
\begin{tabular}{lllllcl}
\br
Experiment&Search Reaction&$\theta^+$&$\Xi_5$&$\theta_c$&Reference\\
\mr
ALEPH&Hadronic Z decays&$\Downarrow $&$\Downarrow $&$\Downarrow $&\cite{aleph}\\
BaBar&$e^+e^- \to \Upsilon(4S)$&$\Downarrow $&$\Downarrow $&$ - $&\cite{babar}\\
BELLE&$KN \to PX$&$\Downarrow $&$-$&$\Downarrow $& \cite{belle}\\
BES&$e^+e^- \to J/\psi (\psi(2S) \to \theta \bar \theta$&$\Downarrow $&$-$&$\Downarrow $& \cite{bes}\\
CDF&$p \bar p \to PX$&$\Downarrow $&$\Downarrow $&$\Downarrow $& \cite{cdf}\\
COMPASS&$\mu^+ (^6LiD) \to PX$&$\Downarrow $&$\Downarrow $&$- $& \cite{compass}\\
DELPHI&Hadronic Z decays&$\Downarrow $&$- $&$- $ &\cite{lep2}\\
E690&$pp \to PX$&$\Downarrow $&$\Downarrow $&$- $&\cite{e690}\\
FOCUS&$\gamma p \to PX$&$\Downarrow $&$\Downarrow $&$\Downarrow $ &\cite{focus}\\
HERA-B&$pA \to PX$&$\Downarrow $&$\Downarrow $&$- $&\cite{herab}\\
HyperCP&$(\pi^+,K^+,p)Cu \to PX$&$\Downarrow $&$- $&$- $ &\cite{hypercp}\\
LASS&$K^+p \to K^+n\pi^+$&$\Downarrow $&$- $&$- $ &\cite{lass}\\
L3&$\gamma \gamma \to \theta \bar \theta$&$\Downarrow $&$- $&$- $ &\cite{lep2,lep1}\\
PHENIX&$Au Au \to PX$&$\Downarrow $&$-$&$- $&\cite{phenix}\\
SELEX&$(\pi, p, \Sigma)p \to PX$&$\Downarrow $&$- $&$- $ &\cite{selex}\\
SPHINX&$p C(N) \to \theta^+ C(N)$&$\Downarrow $&$- $&$- $ &\cite{sphinx}\\
WA89&$\Sigma^-N \to PX$&$- $&$\Downarrow $&$- $& \cite{wa89}\\
ZEUS&$ep \to PX$&$\Uparrow $&$\Downarrow $&$\Downarrow $& \cite{zeus,cascade,zeus_charm}\\
\br
\end{tabular}
\end{center}
\end{table}

The ALEPH, DELPHI and L3 experiments ran at LEP.  The ALEPH \cite{aleph} and
DELPHI \cite{lep2} searches studied hadronic $Z$ decays while L3 \cite{lep2,lep1}
searched for the $\theta^+$ in photon-photon collisions.  The BaBar, BELLE and
BES experiments are also all carried out at $e^+e^-$ colliders.  BaBar searched \cite{babar}
in final states from $e^+e^- \to \Upsilon(4S)$ as well as 40~MeV below resonance
while BES \cite{bes} searched
in $e^+e^- \to J/\psi(\psi(2S)) \to \theta \bar \theta$.  And BELLE searched for 
$\theta^+$ in kaon interactions in the detector material \cite{belle} -- a search
that yielded 16000~$\Lambda(1520) \to pK^-$ decays but no $\theta^+$ signal.
The CDF experiment at FNAL searched for all three pentaquark states \cite{cdf} in
$\bar p p$ collisions.  The E690 \cite{e690}, FOCUS \cite{focus}, HyperCP \cite{hypercp}
 and SELEX \cite{selex} searches were all
carried out in fixed target experiments at FNAL.  FOCUS used a photon beam
produced by bremsstrahlung of 300~GeV electrons and positrons,
E690 a 800~GeV/$c$ proton beam, HyperCP a mixed beam of $\pi^+$, $K^+$ and protons
ranging in momentum from 50 to 250~GeV/$c$
and SELEX a 600~GeV/$c$ mixed beam of $\pi$, $K$ and $\Sigma^-$.  HERA-B \cite{herab} and
SPHINX \cite{sphinx} at HERA and IHEP (Protvino, Russia) also were fixed target
experiments, both using proton beams -- 900~GeV/$c$  at
HERA and 70~GeV/$c$ at IHEP.  
COMPASS \cite{compass} is a 
fixed target experiment at CERN using a 160~GeV/$c$  $\mu^+$ beam.  The LASS spectrometer
at SLAC collected data using a 11~GeV/$c$ $K^+$ beam and these data 
were recently re-analyzed \cite{lass}.  WA89 is a fixed target at experiment
that used a 340~GeV/$c$ $\Sigma^-$ beam at CERN \cite{wa89}.  The
PHENIX detector searched for the $\theta^+$ in $Au-Au$ collisions \cite{phenix}  at RHIC.

\section{Comparing the positive and negative results}

Since the $\Xi_5$ and $\theta_c$ have only been observed by one experiment each,
we discuss these states first and then discuss the $\theta^+$.

\subsection{The $\Xi_5$ and $\theta_c$}

The $\Xi_5$ has only been observed by NA49 \cite{na49} and searched for, but not observed,
by 9 experiments \cite{zeus,aleph,babar,cdf,compass,e690,focus,herab,wa89}
as shown in Table~\ref{negex}.
In the $\Xi^- \pi^-$ mode NA49 reports 38 signal  ($s$)  events over a background
($b$) of 43 events and thus claim a signal significance ($s.d.$) of 4.2~$\sigma$ assuming
$s.d.=s/\sqrt{s+b}$.  When they combine events from the $\Xi^- \pi^+$ mode
as well then $s=69$, $b=75$ and $s.d.=5.8\sigma$.  NA49 observes a total
of 1640~$\Xi^-(1320)$ in their sample.  By contrast ALEPH, HERA-B,
CDF, BaBar and WA89 have samples that exceed the NA49 sample of
$\Xi(1320)$ by factors of 2, 10, 22, 157 and 412 respectively and all but NA49 fail to 
observe the  $\Xi_5$. 
Perhaps even more relevant are the numbers of $\Xi(1530)$ states observed.
NA49 finds 150 $\Xi(1530)$'s \cite{fischer}  while ALEPH and ZEUS
also see a comparable number but CDF sees 6 times more and
BaBar and E690 see 100 times more.
Not only are the statistics of the experiments that do not observe the $\Xi_5$
significantly higher, but the typical mass resolution of 1 to 2~MeV/$c^2$
is much better than the resolution of NA49.
 It is hard to reconcile the existence of the $\Xi_5$
with this overwhelming negative evidence.

The $\theta_c$ has only been observed by H1 \cite{h1} and searched for, but not observed,
by 6 experiments \cite{aleph,belle,bes,cdf,focus,zeus} as shown in Table~\ref{negex},
H1 sees $s \approx 50$,
$b \approx 50$ and claim $s.d. = 5 \sigma$.  The $\theta_c$ is observed in its
decay to $D^{*+} \bar p$
and $D^{*-}  p$.  H1 has $\approx 3000$ $D^*$'s in their sample and by contrast the
number of $D^*$'s observed by FOCUS and CDF exceed the H1 sample by
factors of 12 and 178 respectively.  The failure of ZEUS to find the $\theta_c$ 
is particularly relevant because H1 and ZEUS are similar experiments operating
at the same accelerator facility.
 It is hard to reconcile the existence of the $\theta_c$
with this overwhelming negative evidence.

\subsection{Comparing positive $\theta^+$ sightings with the null results }

As shown in Table~\ref{posex}, the $\theta^+$ has been observed in 13 reactions
where two of the reports are from the LEPS group \cite{spring8,spring8-2} and
another two from the CLAS group \cite{clas1,clas2}.  As shown in Table~\ref{negex},
the $\theta^+$ has been searched for by 16 separate experiments and all those
came up with null results.  

An interesting benchmark to compare the positive sightings with the null results is
the number of observed $\Lambda(1520) \to p K^-$ decays.  The number
of claimed signal events $s_{\theta}$ and observed $\Lambda(1520)$ decays
$s_{\Lambda}$ for LEPS(1) is $s_{\theta}(s_{\Lambda})=$~19(25).  For the other positive 
sightings, $s_{\theta}(s_{\Lambda})$ for LEPS(2), CLAS(d), SAPHIR, HERMES and
ZEUS are 56(162), 43(212), 55(530), 51(850) and 230(193) respectively.  By
contrast, the number of $\Lambda(1520)$ decays observed by ALEPH, BaBar,
BELLE, CDF, E690, HERA-B and SPHINX (in thousands of events) are
2.8K, 100K, 15.5K, 3.3K, 5K and 23K.  So, for example,
 the BaBar yield of $\Lambda(1520)$ decays
exceeds that of LEPS(1) by a factor of 4000.

It seems difficult to reconcile the positive sightings with the null results since
each of the two set of experiments spans a wide variety of production mechanisms.
And in general the experiments yielding null results have higher statistics and
superior resolution.  

\section{Examining the positive $\theta^+$ sightings in more detail}

Since the existence of a flavor-exotic baryon state would have such an important
impact on our understanding of hadronic physics it is essential that the
positive evidence be examined critically.  In particular, we will discuss

\begin{itemize}
\item[a)]  The statistical significance of the purported signals 
including a discussion of the background estimation;
\item[b)]  The effect of kinematical reflections giving rise to enhancements that
could fluctuate to sharp peaks;
\item[c)]  Other possibilities of generating sharp peaks
including {\it ghost tracks}, kinematic cuts and other mechanisms;
\item[d)]  Effects that could impair the isolation of  exclusive reactions; and
\item[e)]  Problems of width estimates and consistency of masses among experiments.
\end{itemize}

\begin{table}[h]
\caption[]{\label{statistics}A tabulation of statistics for the observations
of the $\theta^+$. See text for descriptions of the statistical significance
as quoted in the three columns of ratios. The column labeled Published  is
the significance quoted in the publication. }
\begin{center}
\begin{tabular}{lcccccc}
\br
Experiment & Signal & Background & \multicolumn{4}{ c }{Significance}  \\
           &  $s$   &   $b$      & Published &
$\frac{s}{\sqrt{b}}$ & $\frac{s}{\sqrt{s+b}}$ & 
$\frac{s}{\sqrt{s+2b}}$ \\ 
\mr
LEPS(1)~\cite{spring8}    & $19$ & $17$ & $4.6$     & $4.6$ & $3.2$ & $2.6$  \\
LEPS(2)~\cite{spring8-2}    & $56$ & $162$&           & $4.4$ & $3.8$ & $2.9$  \\
CLAS(d)~\cite{clas1} & $43$ & $54$ & $5.2$     & $5.9$ & $4.4$ & $3.5$  \\
CLAS(p)~\cite{clas2}& $41$ & $35$ & $7.8$     & $6.9$ & $4.7$ & $3.9$   \\
SAPHIR~\cite{saphir}   & $55$ & $56$ & $4.8$     & $7.3$ & $5.2$ & $4.3$  \\
COSY~\cite{cosy}     & $57$ & $95$ & $4-6$     & $5.9$ & $4.7$ & $3.7$  \\
JINR~\cite{jinr}     & $88$ & $192$ & $5.5$     & $6.4$ & $5.3$ & $4.1$  \\
SVD~\cite{svd}      & $35$ & $93$ & $5.6$     & $3.6$ & $3.1$ & $2.4$  \\
DIANA~\cite{diana}   & $29$ & $44$ & $4.4$     & $4.4$ & $3.4$ & $2.7$  \\
$\nu$BC~\cite{bebc}&$18$&$9$& $6.7$     & $6.0$ & $3.5$ & $3.0$  \\
NOMAD~\cite{nomad}&$33$ & $59$ & $4.3$     & $4.3$ & $3.4$ & $2.7$  \\
HERMES~\cite{hermes} & $51$ & $150$& $4.3-6.2$ & $4.2$ & $3.6$ & $2.7$  \\
ZEUS~\cite{zeus}  & $230$&$1080$& $4.6$     & $7.0$ & $6.4$ & $4.7$  \\
\br
%
\end{tabular}
\end{center}
\end{table}

\subsection{Statistical significance of positive sightings}

Table~\ref{statistics} lists the number of signal events and background events
for each of the positive sightings along with the published statistical significance
of the $\theta^+$.   The simple 
estimators rely on two numbers. The first is the number of 
background counts under the peak, $b$, while the second is 
the number of signal counts in the peak, $s$ above background. From these two
numbers, there are three commonly used significance estimators.
\begin{enumerate}
\item[(i)]$ s/\sqrt{b}$
\item[(ii)] $s/\sqrt{s+b}$
\item[(iii)]$ s/\sqrt{s+2b}$
\end{enumerate}
The first of these neglects the statistical uncertainty of the 
background and is normally an over-estimate of the true significance. 
This is  the most commonly quoted estimator for
observed pentaquark signals. The second estimator assumes a smooth 
background with a well defined shape. Finally, the third method takes 
into account the uncertainty in a statistically independent background
under the signal. This latter method should normally underestimate the 
true significance of an observed signal. All three estimators are computed
for each sighting in Table~\ref{statistics}.  Since all of these experiments report 
on a signal with an unknown production
mechanism over a background that is not understood, the best statistical measure
is likely to be case (iii), $s/\sqrt{s+2b}$.

\begin{figure}
\begin{center}
\includegraphics[width=36pc]{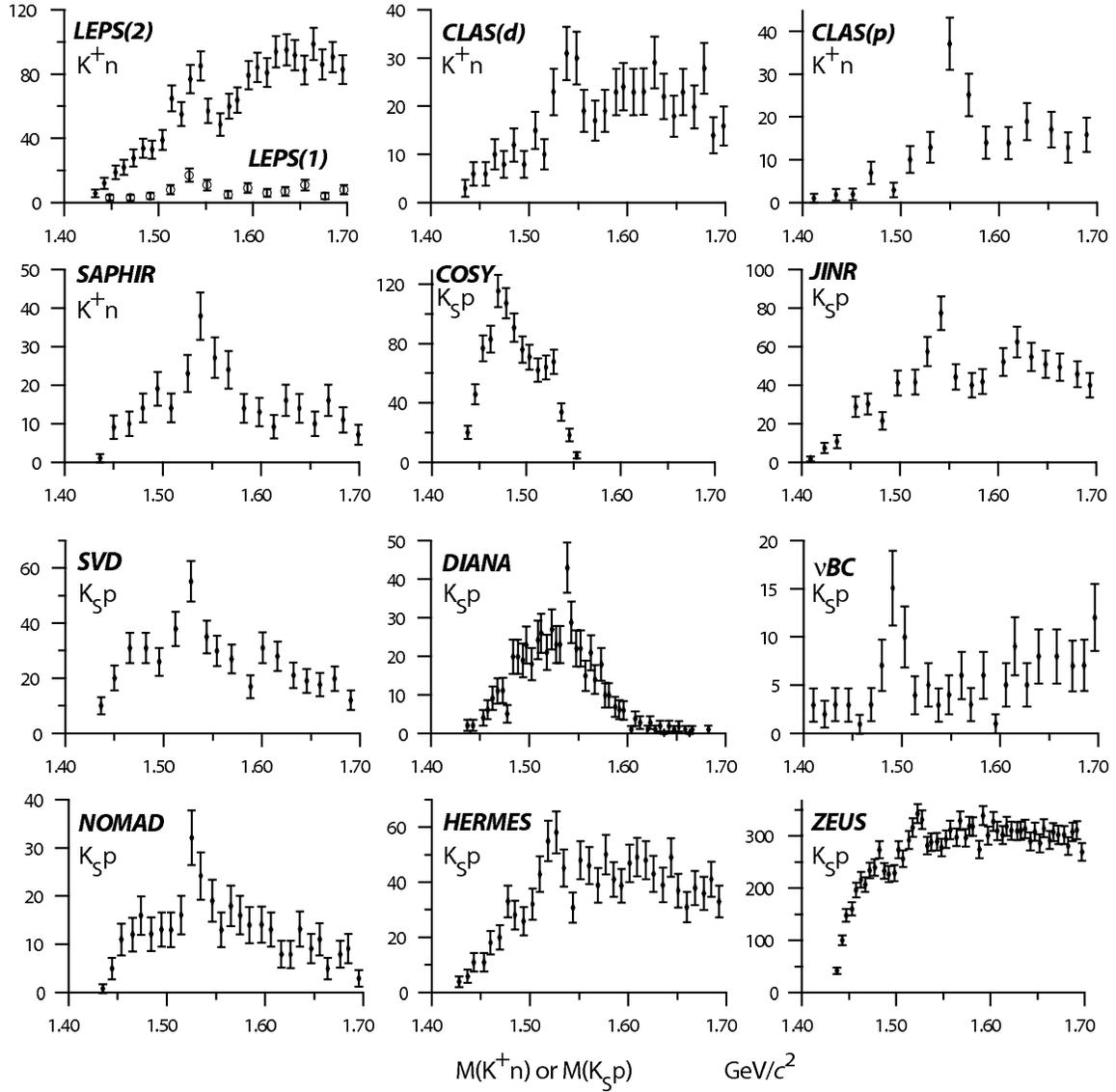}
\end{center}
\caption{\label{allplots}A comparison of the mass distributions for the $K^+n$
or  $K^0_Sp$ mass distributions for the experiments listed in Table~\ref{posex}.
Only the mass region from 1.4 to 1.7~GeV/$c^2$ is shown.  Error bars are 
based simply on statistical errors.}
\end{figure}

In Figure~\ref{allplots} we show  the 
 $K^+n$ or  $K^0_Sp$
mass distributions for the experiments listed in Table~\ref{posex}
absent any background curves.
Only the mass region from 1.4 to 1.7~GeV/$c^2$ is shown.  Error bars are 
based simply on statistical errors.

\begin{figure}[h]
\includegraphics[width=18pc]{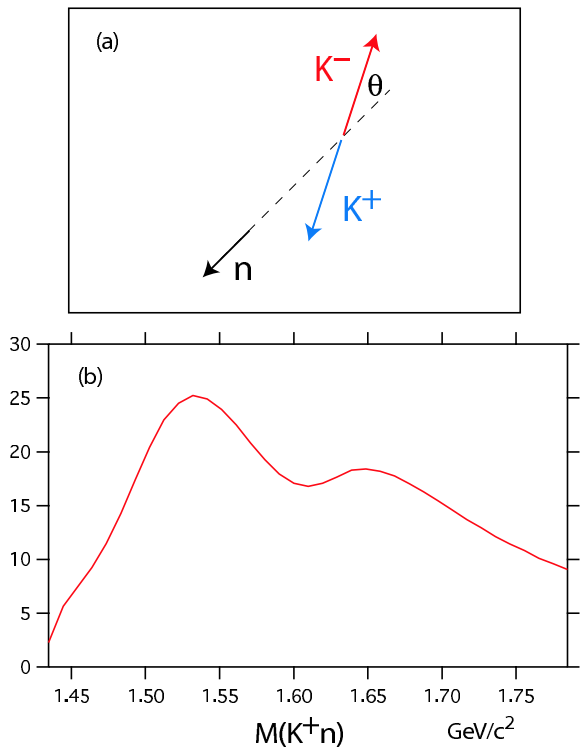}\hspace{2pc}%
\begin{minipage}[b]{18pc}\caption{\label{parent}For the reaction
$\gamma n \to K^+ K^-n$  the configuration of momentum vectors
is shown in Figure~(a) in the $K^+K^-$ rest frame where the boost direction
to this frame defines the axis.  As discussed in \cite{ardaps} the decays
of mesons, such as the $\phi$, $a_2$, $f_2$ and $\rho_3$, with their
characteristic decay structure and interference with each other, will influence
the shape of the $K^+n$ (and $K^+n$ ) mass distribution.  The curve shown in
Figure~(b) is a possible  description of the background for the 
observed $K^+n$ spectrum reported in \cite{clas1} and results from a
simultaneous fit of the $K^+n$, $K^-n$ and $K^+K^-$ effective mass 
distributions with an assumption that the  $\phi$, $a_2$, $f_2$ and $\rho_3$
are produced and then decay into $K^+K^-$.}
\end{minipage}
\end{figure}

\subsection{Kinematic reflections}

It was pointed out in reference~\cite{ardaps} that for lower energy photon experiments
\cite{spring8,spring8-2,clas1,clas2,saphir}
the production of mesons, such as the $\phi$, $a_2$, $f_2$ and $\rho_3$, with
subsequent decays into $K^+K^-$ could lead to structure in the the $K^+n$ mass
distribution.  This comes about because of the decay pattern of the decaying mesons
given by $|Y^m_{\ell}(\theta,\phi)|^2$ and is illustrated in Figure~\ref{parent}.  In
part (a) of Figure~\ref{parent} the momentum vectors in the rest frame of the decaying
mesons is shown.  The boost direction is along the line-of-flight of the meson 
and in this {\it helicity frame} that direction defines
 the $z-$axis.  When the polar angle $\theta$
is small, the $K^+$ and $n$ momenta are nearly collinear and the  $K^+n$ effective mass
is small.  For $\theta \approx \pi$ the effective mass is large.  There is strong 
forward-backward peaking because of the $|Y^m_{\ell}(\theta,\phi)|^2$ and this
reflects in the $K^+n$ effective mass spectrum.  But because of the constrained
kinematics, owing to the low beam momenta, only the lower peak survives.
 The curve shown in
Figure~\ref{parent}(b) is a possible  description of the background for the 
observed $K^+n$ spectrum reported in \cite{clas1} and results from a
simultaneous fit of the $K^+n$, $K^-n$ and $K^+K^-$ effective mass 
distributions with an assumption that the  $\phi$, $a_2$, $f_2$ and $\rho_3$
are produced and then decay into $K^+K^-$.

In 1969 Anderson {\it et al} \cite{anderson} performed a search
for the $Z^*$ in $\pi^- p \to K^- Z^*$ at 6 and 8~GeV/$c$ by looking
at the missing mass recoiling against the $K^-$ (as is the case as
well for the LEPS(1), LEPS(2) and CLAS(d) searches).
 They found a
peak at around 1.6~GeV/$c^2$ that they eventually ascribed to a
kinematic reflection due to production of mesons decaying into $K \bar K$.  
This was supported by Monte Carlo studies and the
observations that the peak position changed as the beam energy
changed.

\begin{figure}[h]
\includegraphics[width=22pc]{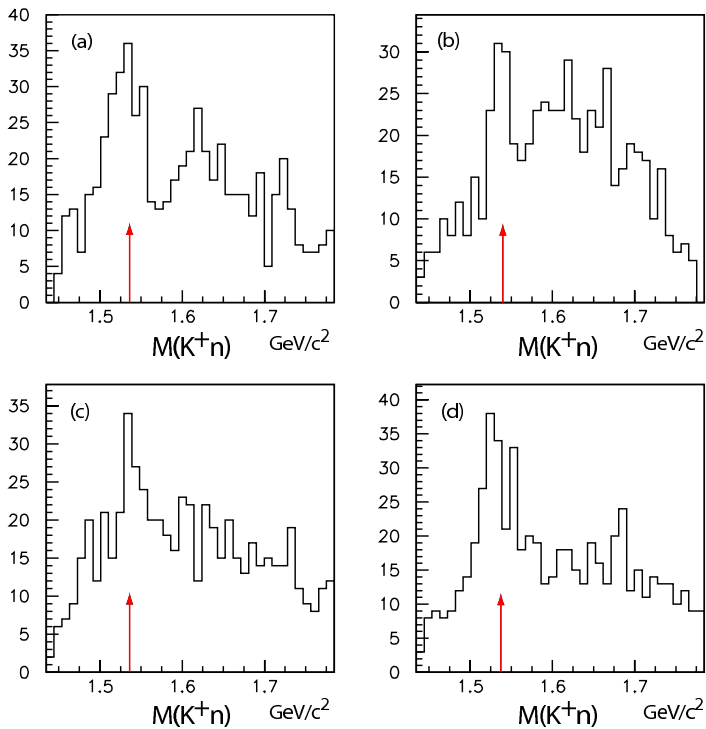}\hspace{2pc}%
\begin{minipage}[b]{14pc}\caption{\label{fakes}Using as a parent distribution, 
the distribution shown in Figure~\ref{parent}, 20 histograms were generated
with statistics equal to the $K^+n$ distribution presented in 
\cite{clas1} -- also shown in Figure~(b).  Of the 20 histograms, three were selected that had
a sharp peak near the mass of $\theta^+$.  These are shown in
Figures ~(a), (c) and (d).}
\end{minipage}
\end{figure}

Although such a mechanism cannot produce a sharp peak in the $K^+n$
mass distribution it will reduce the statistical significance of a purported
signal in the low-mass region where this background peaks.  And it could
well lead to fluctuations into a sharp peak with low statistics as is illustrated
in Figure~\ref{fakes}.

Using as a parent distribution
the distribution shown in Figure~\ref{parent}, twenty histograms were
randomly  generated
with statistics equal to the $K^+n$ distribution presented in 
\cite{clas1} -- also shown in part~(b) of Figure~\ref{parent}. 
 Of the 20 histograms generated,  three were selected that had
a sharp peak near the mass of $\theta^+$.  These are shown in
parts~(a), (c) and (d) of Figure~\ref{fakes}.  The use of this mechanism
to estimate the background for the CLAS(d) \cite{clas1} data would
result in a 20\% probability of fluctuating into a sharp peak at the
$\theta^+$ mass.

\subsection{Spurious peaks}

Of the 13 $\theta^+$ sightings list in Table~\ref{posex}, 8 are of the decay $\theta^+ \to K_S^0p$
and of those all but two, COSY and DIANA, are inclusive measurements.  It was pointed
out by Zavertyaev \cite{zavertyaev} and
 Longo  \cite{longo} that ghost tracks associated with the decay $\Lambda(1115) \to \pi^- p$
could lead to a spurious sharp spike at precisely 1.54~GeV/$c^2$
if $p_{\Lambda} >$~2~GeV/$c$.  The inclusive 
experiments should produce $\Lambda$'s copiously.  

We illustrate this mechanism in Figure~\ref{fakes}.  In part~(a) of that figure we show
a schematic of a $\Lambda(1115) \to \pi^- p$ decay where the track reconstruction
duplicated the proton track yielding an extra spurious positively charged track
that has been assigned the mass of a pion.  
When
combined with the $\pi^-$ from the $\Lambda^0$ the effective mass clusters about 
0.5~GeV/$c^2$ as in part~(b) and when the ghost track is combined with the
$\Lambda^0$ decay products the effective mass clusters around 1.5~GeV/$c^2$
as seen in part~(c).  
In the shaded distributions the $"\pi^+"\pi^-$ mass is required to be near the
$K_S^0$.  The mean of the shaded portion of the distribution in part~(c) is
1.54~GeV/$c^2$, the mass of the $\theta^+$.  In this study the $\Lambda^0$ momentum
in the LAB frame was uniform from 2 to 100~GeV/$c$.

The DIANA experiment shows a peak at the $\theta^+$ mass of 1.54~GeV/$c^2$
(see Figure~\ref{allplots}).  But as pointed out by Zavertyaev \cite{zavertyaev}, among
the observed reactions should be charge exchange off a neutron, $K^+n \to K_S^0p$.
This two-body reaction should yield a fixed $K_S^0p$ effective mass
if the $K^+$ beam momentum is fixed.  Indeed the beam momentum spectrum is consistent with the
position of the observed peak in the DIANA spectrum.

\begin{figure}[h]
\includegraphics[width=22pc]{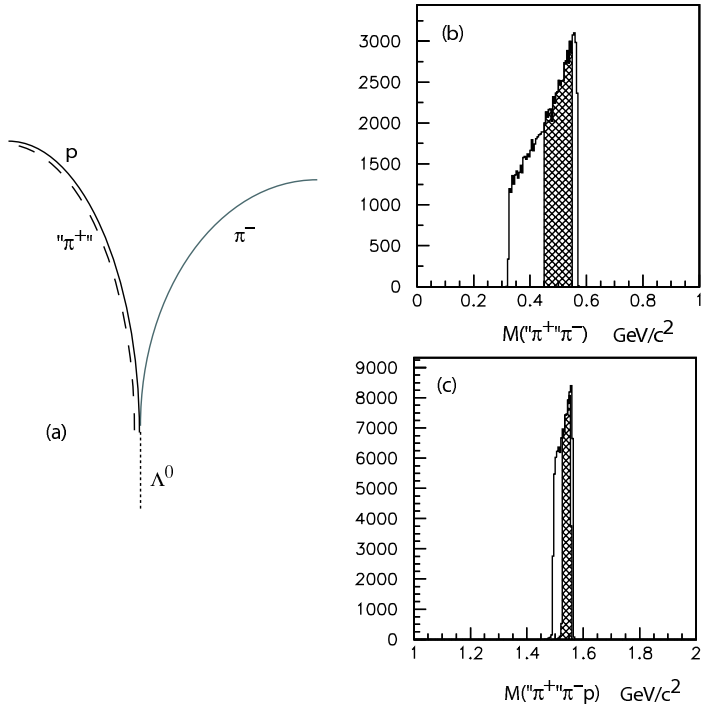}\hspace{2pc}%
\begin{minipage}[b]{14pc}\caption{\label{p1}Figure (a) is a schematic of the decay
$\Lambda^0(1115) \to \pi^- p$.  The effect of spurious \emph{ghost} tracks from the 
reconstruction software is considered.  In this case a $\pi^+$ track is generated.  When
combined with the $\pi^-$ from the $\Lambda^0$ the effective mass clusters about 
0.5~GeV/$c^2$ as in Figure~(b) and when the ghost track is combined with the
$\Lambda^0$ decay products the effective mass clusters around 1.5~GeV/$c^2$
as seen in Figure~(c).  
In the shaded distributions the $"\pi^+"\pi^-$ mass is required to be near the
$K_S^0$.  The mean of the shaded portion of the distribution in Figure~(c) is
1.54~GeV/$c^2$, the mass of the $\theta^+$.  In this study the $\Lambda^0$ momentum
in the LAB frame was uniform from 2 to 100~GeV/$c$.}
\end{minipage}
\end{figure}

The CLAS(p) reaction $\gamma p \to K^+K^- \pi^+ n$ \cite{clas2} is an exclusive reaction
uncomplicated by nuclear target effects (see the next subsection).  After cuts are
made to isolate this reaction the published $K^+n$ mass spectrum shows no structure.
But then kinematic cuts are made to remove processes that are potential
backgrounds to $\theta^+$ production such as:

\begin{enumerate}
\item[(i)]$ \gamma p \to X^+ n$ followed by $X^+ \to K^+ K^{*0} \to K^+ K^- \pi^+$
\item[(ii)] $\gamma p \to X^0 N^{*+}/\Delta^+$ followed by $X^0 \to K^+K^-$ and  $N^{*+}/\Delta^+ \to n \pi^+$
\item[(iii)]$\gamma p \to (X^0 \pi^+)_{forward}n$ followed by $X^0 \to K^+K^-$
\end{enumerate}

To remove these "backgrounds" the following cuts are made:  
$\left| t_{\gamma \to \pi} \right| \le 0.28$~(GeV/$c$)$^{2}$ and $\cos \theta^*_{K^+} \le 0.6$
where $\theta^*_{K^+}$ is the angle of the $K^+$ in the overall center-of-mass.  The result
is a $K^+n$ mass distribution sharply reduced in statistics but with a peak at the $\theta^+$.  
Furthermore it is claimed that the $K^- \theta^+$ effective mass shows a peak at 2.4~GeV/$c^2$
leading to the conclusion that an important mode for $\theta^+$ production is
$\gamma p \to \pi^+ N^*(2400)$ followed by $N^*(2400) \to K^- \theta^+$.  The authors
do not discuss the stability of observed peaks with respect to the kinematic cuts.  We
did Monte Carlo studies of reactions such as $\gamma p \to a_2^0 \Delta^+$ 
and found that the kinematic cuts can produce peaks observed in the CLAS(p) data
depending on details of the spin-state of the $a_2$.  

Stoler also pointed out \cite{stoler} that the these data do show an
anti-correlation of $\theta^+$ signal with the $K^*$ signal which is surprising since
one might expect associated production of the $\theta^+$, {\it i.e.} 
$\gamma p \to K^{*0} \theta^+$ to be favored.  
However, this anti-coorelation would be expected from a kinematical
reflection.

\subsection{Kinematic identification of reactions}


The LEPS(1) and LEPS(2) reactions take place off a carbon nucleus 
leading to complications due to Fermi motion of the target nucleon. 
Because only the  $K^+$ and $K^-$ are measured, they then must assume that the unobserved
neutron carries off the remaining missing momentum. In the case of CLAS(d),
the $\theta^+$ is observed in the effective mass of $K^+n$ where the
$n$ is missing.
 However, to be able to reconstruct the 
reaction, the proton is required to have sufficient energy to escape the 
target. The proton is required to somehow be involved in the reaction, which
leads to complications in understanding both the reaction itself and the
backgrounds to the reaction. 

The COSY experiment \cite{cosy} depends on identification of the reaction
$pp \to \Sigma^+ K_S^0p$ but this experiment is highly unconstrained.  
This only information about this experiment is a collection of space-points
from three pairs of hodoscope planes downstream of a "point" target.  There is no
momentum measurement, time measurement nor particle identification.   The $K_S^0$ 
is identified through its decay into $\pi^+ \pi^-$ between the first and second
hodoscope planes and the $\Sigma$ is identified as a 'kink'.  This experiment
is seriously under-constrained and subject to backgrounds.

\begin{figure}
\begin{center}
\includegraphics[width=20pc]{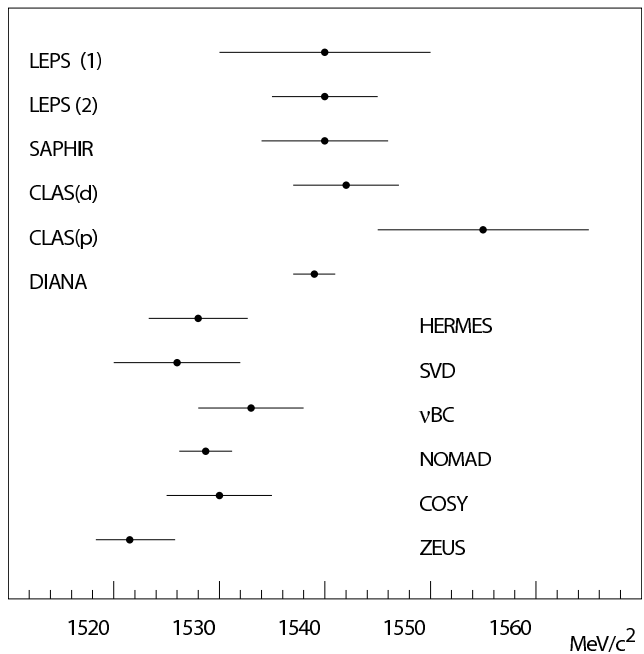}
\end{center}
\caption{\label{masses2}Reported masses, with error bars, of the $\theta^+$.}
\end{figure}

\subsection{The reported masses and widths of the $\theta^+$}

Figure~\ref{masses2} shows a plot of the masses reported for the $\theta^+$.  The quoted
masses are inconsistent with each other although the reported widths are less 
than experimental resolution and these experiments accurately 
reproduce masses of established resonances.

Cahn and Trilling \cite{trilling_cahn} analyzed the data from the DIANA experiment
and conclude that the width of the $\theta^+$ has a width of less than 1~MeV/$c^2$.
Arndt  {\it et al} \cite{arndt} reanalyzed $K^+N$ and conclude that the addition of the
$\theta^+$ with a width of about 5~MeV/$c^2$ is improbable but the data are consistent with a width of the
$\theta^+$ less than  1~MeV/$c^2$.

\section{Summary and conclusions} 

The experimental evidence for the pentaquark states $\theta^+$, $\Xi_5$ and $\theta_c$
has been reviewed.
The $\theta^+$ has been observed in 13 reactions
where two of the reports are from the LEPS group \cite{spring8,spring8-2} and
another two from the CLAS group \cite{clas1,clas2}.  The
$\theta^+$ has been searched for by 16 separate experiments and all those
came up with null results.  The positive sightings are complicated by nuclear target effects, 
mechanisms that can generate spurious peaks and problems with kinematic identification
of final states.  Furthermore, stringent cuts, the stability of which are not well understood, are
required to isolate the peaks.  On the other hand, experiments that have searched for
the $\theta^+$ with null results have high statistics and superior resolution. 
The 
reported masses for the $\theta^+$ are inconsistent with each other and the width
is unusually small for a typical hadronic state. 
There have been attempts, {\it e.g.} Karliner and Lipkin
\cite{kali}, to understand all these results
in terms of models that would suppress pentaquark production in some processes
and not others.  If it is real, the $\theta^+$ is exotic 
not only in its flavor quantum numbers but also in its production and its decay.

The $\Xi_5$ has only been observed by NA49 \cite{na49} and searched for, but not observed,
by 9 experiments \cite{zeus,aleph,babar,cdf,compass,e690,focus,herab,wa89}.
 It is hard to reconcile the existence of the $\Xi_5$
with this overwhelming negative evidence.

The $\theta_c$ has only been observed by H1 \cite{h1} and searched for, but not observed,
by 6 experiments \cite{zeus,aleph,belle,bes,cdf,focus}.
 It is hard to reconcile the existence of the $\theta_c$
with this overwhelming negative evidence.

Based on all this, the conclusion is that the experimental evidence for pentaquarks
is very weak.

\section{Acknowledgments}

The authors wish to acknowledge helpful discussions with D.~Christian, M.~Longo, 
R.~Mitchell, M.~Reyes and S.~Teige.  This work was supported in part by the Department
of Energy.

 \end{document}